**Electron carrier-mediated room temperature ferromagnetism in anatase $(Ti,Co)O_2$**

Tomoteru Fukumura

Department of Chemistry, University of Tokyo, Tokyo 113-0033, Japan

Yoshinori Yamada[a]

Quantum Phase Electronics Center and Department of Applied Physics, University of Tokyo, Tokyo 113-8656, Japan

Kazunori Ueno

Department of Basic Science, University of Tokyo, Tokyo 153-8902, Japan and PRESTO, Japan Science and Technology Agency, Kawaguchi 332-0012, Japan

Hongtao Yuan[b]

Quantum Phase Electronics Center and Department of Applied Physics, University of Tokyo, Tokyo 113-8656, Japan

Hidekazu Shimotani

Department of Physics, Tohoku University, Sendai 980-8578, Japan

Yoshihiro Iwasa

Quantum Phase Electronics Center and Department of Applied Physics, University of Tokyo, Tokyo 113-8656, Japan

Lin Gu[c]

WPI-AIM Research, Tohoku University, Sendai 980-8577, Japan

Susumu Tsukimoto

WPI-AIM Research, Tohoku University, Sendai 980-8577, Japan

---

[a] On leave from Tohoku University, Japan

[b] Present address: Stanford University, USA

[c] Present address: Chinese Academy of Science, China

Yuichi Ikuhara

WPI-AIM Research, Tohoku University, Sendai 980-8577, Japan, Institute of Engineering Innovation, University of Tokyo, Tokyo 113-8656, Japan, and Nanostructures Research Laboratory, Japan Fine Ceramics Center, Nagoya 456-8587, Japan

Masashi Kawasaki

Quantum Phase Electronics Center and Department of Applied Physics, University of Tokyo, Tokyo 113-8656, Japan

Since the discovery of room temperature ferromagnetism in (Ti,Co)$O_2$, the mechanism has been under discussion for a decade. Particularly, the central concern has been whether or not the ferromagnetic exchange interaction is mediated by charge carriers like (Ga,Mn)As. Recent two studies on the control of ferromagnetism in anatase (Ti,Co)$O_2$ at room temperature via electric field effect [Y. Yamada et al., Science **332**, 1065 (2011)] and chemical doping [Y. Yamada et al., Appl. Phys. Lett. **99**, 242502 (2011)] indicate a principal role of electrons in the carrier-mediated exchange interaction. In this article, the authors review fundamental properties of anatase (Ti,Co)$O_2$ and discuss the carrier mediated ferromagnetism.

## 1. Introduction

Diluted magnetic semiconductor (DMS) is a semiconductor doped with a transition metal element, possessing both semiconducting and magnetic characters. One of the well-known DMS is Mn-doped II−VI compound semiconductors that are usually not ferromagnetic [1]. Low temperature molecular beam epitaxy enables to grow ferromagnetic Mn-doped III-V compound semiconductors [2], what is called ferromagnetic semiconductors, the Curie temperature ($T_C$) of which is as high as 190 K [3]. In these DMS, the band carriers mediate exchange interaction between localized spins of the transition metal elements via *sp*−*d* exchange interaction. The ferromagnetism in (Ga,Mn)As is useful to demonstrate semiconductor spintronics [4,5], in spite of difficulty in the device operation at room temperature.

Oxide semiconductors such as ZnO are expected to serve as a host of higher $T_C$ ferromagnetic semiconductor, because their heavier carrier mass, owing to their wide gap, as well as heavy carrier doping will contribute to strong carrier-mediated exchange interaction such as Ruderman−Kittel−Kasuya−Yosida interaction [6]. Possibilities of higher $T_C$ was also proposed from theoretical sides [7,8]. (Zn,Mn)O was reported as a wide gap magnetic oxide semiconductor exhibiting quite similar properties to the ordinary II-VI DMS [6,9], where the smallest oxide anion among chalcogenide ions led to stronger antiferromagnetic exchange interaction [9,10]. Subsequently, new oxide-based DMS such as 3*d* transition metal (Sc−Cu)-doped ZnO [11] and $(Sn,Mn)O_2$ [12] were synthesized. And finally, room temperature ferromagnetism was discovered in $(Ti,Co)O_2$ with both anatase and rutile structures [13,14]. Figure 1 shows the magnetic hysteresis at room temperature and the temperature dependence of magnetization for $(Ti,Co)O_2$ grown in early stage.[d] The $T_C$ is much higher than those of representative ferromagnetic semiconductors: (Zn,Cr)Te, (In,Mn)As, and (Ga,Mn)As [15−17].

Until now, a large number of studies postulated high temperature ferromagnetism in doped oxides [18−24]. The most concern was the origin of the ferromagnetism: either carrier-mediated mechanism, defect-mediated mechanism, or extrinsic effect [25,26]. Recent studies, the control of ferromagnetism by electric field [27] and chemical carrier doping [28], clarified a principal role of electron carriers to mediate the ferromagnetism in Co-doped anatase $TiO_2$. In this article, the authors review basic properties of this system in Section 2, the electric field effect of the ferromagnetism in Section 3, the magnetic phase diagram in Section 4, and summary of these results in Section 5. Fundamental properties of rutile $(Ti,Co)O_2$ are described elsewhere [29].

## 2. Basic properties

### 2.1. Epitaxial thin film growth

Thin film growth of $(Ti,Co)O_2$ was carefully optimized in order to prevent any segregation of metallic Co clusters in such thermodynamically nonequilibrium phase. Ultrahigh vacuum pulsed laser deposition was used for epitaxial growth of anatase $(Ti,Co)O_2$ thin film on atomically flat $LaAlO_3$ substrate buffered with five monolayer-thick insulating $TiO_2$ (Fig. 2(a)) [27]. The buffer layer enabled the low temperature growth of $(Ti,Co)O_2$ thin film below 300°C [30]. Atomic step and terrace structure of the film surface was observed with atomic force microscope. Uniform contrast of Co inside the film without segregation at surface/interface was confirmed by the energy-dispersive x-ray spectroscopy mapping (Fig. 2(b)). Also, the coherent growth of $(Ti,Co)O_2$ thin film and the $TiO_2$ buffer layer was confirmed by the atomically resolved high-angle annular dark field scanning transmission electron microscope (HAADF-STEM) image (Figs. 2(c)−(e)). Ionic state of Co in $(Ti,Co)O_2$ was evaluated to be divalent from x-ray absorption spectroscopy and x-ray magnetic circular dichroism spectroscopy [31], being consistent with recent study from different research group [32].

[d] The narrow hysteresis and the small magnetization were improved later as shown below.

### 2.2. Transport properties

$(Ti,Co)O_2$ becomes electrically conducting by introducing oxygen vacancies as an electron donor as is the case for pure $TiO_2$ [33]. By tuning the oxygen pressure during growth, the amount of oxygen vacancy was finely controlled, so that the carrier density was also fine-tuned. The high quality buffer layer was useful for this reproducible fine-tuning.

Figure 3(a) shows temperature dependence of resistivity for anatase $(Ti,Co)O_2$ grown in different oxygen pressures [28]. The resistivity decreased monotonically with decreasing oxygen pressure, and the hopping conduction turned into metallic conduction, in contrast to the dominance of hopping conduction in rutile $(Ti,Co)O_2$ due to much lower mobility [29]. Figure 3(b) shows the conductivity, electron density, and mobility at 300 K for each sample [28]. The relatively unchanged mobility, from $10^0$ to $10^1$ $cm^2/Vs$, in comparison with the electron density, from $10^{18}$ to over $10^{20}$ $cm^{-3}$, represents that the oxygen vacancy properly served as the electron donor.

### 2.3. Magnetic properties

Magnetization is a fundamental quantity for ferromagnetism, although there can be various misleading factors in magnetometry of thin films due to the small measurement signal. Hence, more direct measurements of ferromagnetic signal due to exchange interaction are vitally important. For example, magnetic circular dichroism (MCD) is a magnetic response of photoexcited band carriers coupled with *d* electrons of transition metal ions via the exchange interaction. This coupling is manifested as the similarity of MCD spectra to the energy derivative of absorption coefficient [34]. Anomalous Hall effect is a magnetic response of itinerant band carriers caused by asymmetric carrier scattering in the presence of spin–orbit interaction, being important fingerprint of the exchange interaction [35]. In addition, x-ray MCD is also useful to element-selectively measure the magnetization signal, ruling out the possibility of unintentional magnetic elements, as was demonstrated for anatase (Ti,Co)$O_2$ [31]. The above measurements are useful criteria of the intrinsic ferromagnetism.

### 2.3.1. Magnetization

Figures 4(a) show magnetic field dependence of magnetization at 300 K for anatase (Ti,Co)$O_2$ with different electron densities [28]. For the lowest electron density, the magnetization was negligible with closed hysteresis, and the magnetization increased monotonically with increasing the electron density accompanied by the enhanced hysteresis. The saturation magnetization was about 2 $\mu_B$/Co, which is larger than that of the rutile (Ti,Co)$O_2$ [29]. Figure 4(b) shows the magnetic field dependence of magnetization at 300 K for anatase (Ti,Co)$O_2$ with different electron densities under out-of-plane and in-plane external fields [27]. The out-of-plane and in-plane magnetizations increased with increasing the electron density, where the out-of-plane magnetization was larger than the in-plane magnetization. Therefore, easy magnetization axis was along the out-of-plane, suggesting that the magnetic anisotropy was not significantly changed in this range of the electron density.

### 2.3.2. Magnetic circular dichroism

Figure 5 shows the absorption and MCD spectra for anatase $TiO_2$ and $(Ti,Co)O_2$ at room temperature [36]. These films were grown on $(La,Sr)_2AlO_4$ substrates instead of $LaAlO_3$ because the twin structures in $LaAlO_3$ hindered the acquisition of intrinsic MCD signal. A slight in-gap absorption in $(Ti,Co)O_2$ is possibly due to the presence of Co *d*-level and/or oxygen vacancy level. Two absorption peaks above the absorption edge correspond to the critical points of the band structure in $TiO_2$ [37] (Fig. 5(a)). As shown in Fig. 5(b), $TiO_2$ showed no MCD signal, except above ~4 eV caused by $LaSrAlO_4$ substrate. On the other hand, $(Ti,Co)O_2$ showed the large negative MCD at visible range and the large positive MCD for higher photon energy with a positive peak around the absorption edge. Magnetic field dependence of MCD signal indicated similar ferromagnetic hysteresis at any photon energy. The MCD spectrum showed two shoulder structures at the critical points of $TiO_2$, indicating the close connection with the host energy band.

### 2.3.3. Anomalous Hall effect

The Hall effect in $(Ti,Co)O_2$ is the sum of the anomalous Hall effect and the ordinary Hall effect, that are connected with the magnetization and the electron density, respectively. The inset of Fig. 6(a) shows the raw Hall resistivity of anatase $(Ti,Co)O_2$ [38]. The ordinary Hall term was dominant over the anomalous Hall term due to the lower electron density than that of rutile $(Ti,Co)O_2$ [29]. The anomalous Hall resistivity $\rho_{AH}$ was deduced by subtracting the ordinary Hall resistivity that is proportional to the magnetic field. Magnetic field dependence of the anomalous Hall resistivity was similar to that of the magnetization (Fig. 6(a)) [38].

The anomalous Hall conductivity $\sigma_{AH}$ is monotonically increasing functions of the magnetization and the conductivity $\sigma_{xx}$, where the latter is proportional to carrier density, hence $\sigma_{AH}$ is more appropriate quantity than the anomalous Hall resistivity to describe the carrier-mediated ferromagnetism in ferromagnetic semiconductors [27]. Figure 6(b) shows magnetic field dependence of the anomalous Hall conductivity at 300 K for anatase $(Ti,Co)O_2$ with different electron densities [28]. The anomalous Hall conductivity was approximately zero for the lowest electron density, and increased with increasing the electron density accompanied by the enhanced hysteresis. This result represents a paramagnetic to ferromagnetic transition, as a manifestation of the carrier-mediated ferromagnetism, being consistent with the result of electric field effect study as described below.

The anomalous Hall conductivity in $(Ti,Co)O_2$ was found to be scaled well with the conductivity, approximately $\sigma_{AH} \propto \sigma_{xx}^{1.6}$ [39], the theoretical background of which was described elsewhere [40]. Figure 7 shows the scaling relation between the anomalous Hall conductivity and the conductivity for various ferromagnetic semiconductors and metals modified from [41], including recent studies [42].

## 3. Electric field effect on ferromagnetism

### 3.1. Electric field effect on ferromagnetic semiconductor

Electric field effect on ferromagnetic semiconductor should enable us to switch on (off) the ferromagnetism by accumulating (depleting) the carriers owing to the low carrier system. In case of ferromagnetic metals, on the other hand, the electric field effect is usually used to control only magnetic anisotropy [43,44], although the switching was recently demonstrated for ultrathin films [45]. The switching of ferromagnetism was firstly demonstrated in (In,Mn)As by accumulating/depleting hole carriers leading to the control of $T_C$ [46]. The magnetization reversal/rotation was also demonstrated in (In,Mn)As and (Ga,Mn)As [47,48].

### 3.2. Electric field induced ferromagnetism in Co-doped $TiO_2$

The electric field effect on ferromagnetism is a powerful tool to clarify the role of carriers in the ferromagnetism because only variable is the amount of carriers. Therefore, it is possible to clarify the question whether or not the oxygen vacancy-mediated interaction without carriers is responsible, as has been discussed in $(Ti,Co)O_2$ [49−51]. In our study [27], electric double layer transistor (EDLT) structure was adopted, where a liquid electrolyte was used in place of solid gate insulator. By using EDLT, the application of high electric field is possible without dielectric breakdown. Also, the liquid electrolyte may avoid magnetoelastic effect, which was caused by piezoelectric material used as a gate insulator [52]. The examination of the Hall effect under electric field is quite useful because the relation between the carrier density and the ferromagnetism can be evaluated from the ordinary and anomalous Hall terms.

Inset of Fig. 8(a) shows a schematic diagram of EDLT. Pt planar gate electrode was located adjacent to the Hall bar. The gate electrode and the $(Ti,Co)O_2$ channel were in contact with the electrolyte. Figure 8(a) shows temperature dependence of resistivity at different gate voltages. With application of positive gate voltages, the insulating behavior in $(Ti,Co)O_2$ turned into metallic conduction. Figure 8(b) shows temperature dependences of the conductivity, mobility, and electron density deduced from the ordinary Hall term at different gate voltages. The mobility was almost constant at different gate voltages, hence the increased conductivity was caused by the increased electron density. The pristine insulating conductivity was recovered after turning off the gate voltage, ruling out the possibility of electrochemical reaction in $(Ti,Co)O_2$ channel.

The variation of magnetization was probed by the anomalous Hall conductivity. Figure 9(a) shows magnetic field dependence of anomalous Hall conductivity at different gate voltages. The anomalous Hall conductivity was nearly negligible at zero gate voltage, corresponding to a paramagnetic state. For higher gate voltages, clear hysteresis emerged, corresponding to the electric field-induced ferromagnetic state at 300 K. Figure 9(b) shows the relation between the anomalous Hall conductivity and the electron density at 300 K for the two EDLTs and for chemically doped $(Ti,Co)O_2$ samples. These all data showed that the anomalous Hall conductivity steeply increases above $n \sim 1 \times 10^{19}$ $cm^{-3}$, corresponding to the paramagnetic to ferromagnetic transition. Good coincidence between the electric field gating and chemical carrier doping in Fig. 9(b) rules out the possibility of any magnetic segregation, and represents that the ferromagnetic exchange interaction in $(Ti,Co)O_2$ is mediated by the electron carriers.

## 4. Magnetic phase diagram via chemical doping

As described above, the carrier-mediated exchange interaction plays a principal role in the ferromagnetism. As shown in Fig. 3(b), the electron density was controlled in a wider range by the chemical doping than that of the electric field effect. As a result, clearer paramagnetic to ferromagnetic transition was observed in the magnetization and the anomalous Hall effect (Figs. 4(a) and 6(b)), in addition to the transformation of insulating to metallic conduction (Fig. 3(a)). Figure 10 summarizes the incremental magnetization from a paramagnetic magnetization of the most insulating sample I, and the anomalous Hall conductivity at 300 K as a function of the electron density [28]. With increasing the carrier density, the magnetization increased at $n$ ~$5 \times 10^{18}$ $cm^{-3}$ (Fig. 10(a)), whereas the anomalous Hall conductivity emerged at $n$ ~$1 \times 10^{19}$ $cm^{-3}$ (Fig. 10(b)). These monotonically increased quantities represent that electron carriers mediate the exchange interaction. Figure 10(b) also shows the data of rutile $(Ti,Co)O_2$ (square symbols) [39]. The rutile $(Ti,Co)O_2$ showed similar tendency to anatase $(Ti,Co)O_2$, but the onset value of the electron density for ferromagnetism was an order of magnitude larger than that of anatase $(Ti,Co)O_2$. This result represents that the rutile $(Ti,Co)O_2$ required much higher electron density to induce the ferromagnetism, probably due to the hopping conduction nature in rutile $(Ti,Co)O_2$ with the heavier electron mass ($m^*$ ~$10\ m_0$ for rutile [53], $m^*$ ~$m_0$ for anatase [54], where $m_0$ is the mass of free electron). It is noted that the sample A in Fig. 10 (open circle symbols) showed closed hysteresis in both magnetization and anomalous Hall conductivity corresponding to a superparamagnetic character, probably caused by the segregation of Co nanoparticles due to excessively reductive growth conditions. This result might be related with previous observation of Co segregations [55,56], implying the improper sample quality.

Let us discuss more details of electron density dependence of the magnetism. In Fig. 10, the low carrier region of $n \leq 5 \times 10^{18}$ $cm^{-3}$ and the high carrier region $n \geq 1 \times 10^{19}$ $cm^{-3}$ correspond to a paramagnetic insulator phase and a ferromagnetic metal phase, respectively. On the other hand, the medium carrier region of $5 \times 10^{18}$ $cm^{-3}$ $< n < 1 \times 10^{19}$ $cm^{-3}$ corresponds to a ferromagnetic insulator phase, in which the anomalous Hall effect is negligible. This result suggests the presence of microscopic ferromagnetic bubble/droplet regions separated by paramagnetic insulator matrix. This kind of phase separation state is present in colossal magnetoresistive manganites and (Ga,Mn)As [3,57−60]. A ferromagnetic insulating phase in $(Ti,Co)O_2$ previously reported [50] could be attributed to such state, probably originated from insufficient carrier density to induce long range ferromagnetic order.

**5. Summary**

The electric field effect and chemical doping studies elucidated a principal role of electron carriers in anatase $(Ti,Co)O_2$. The magnetic and electronic phases in $(Ti,Co)O_2$ were found to vary significantly with the electron density. The Co segregation and the presence of ferromagnetic insulator phase reported previously could be caused by specific sample preparation conditions: the excessively reductive condition and the insufficient carrier doping, respectively. The carrier density dependence of the magnetism unveiled whole picture of the magnetism in $(Ti,Co)O_2$ reflecting the nature of ferromagnetic semiconductors: the essentially inseparable relation between the electronic conduction and the ferromagnetism. The electric field control of ferromagnetism at room temperature will pave the way to room temperature operation of spintronic devices. Nevertheless, there remains an important question: why $T_C$ in $(Ti,Co)O_2$ is so high [61−63]. More quantitative study is needed in order to clarify the mechanism.

**Acknowledgements**

This work was supported in part by JSPS through NEXT Program initiated by CSTP, JST-PRESTO, and MEXT-KAKENHI.

## Figure captions

**FIG. 1.** (a) Magnetic field dependence of magnetization with in-plane magnetic field for an anatase $Ti_{0.93}Co_{0.07}O_2$ thin film in early stage. A photograph of the sample is shown in the inset. (b) Temperature dependence of magnetization for the anatase $Ti_{0.93}Co_{0.07}O_2$ thin film with in-plane magnetic field [13].

**FIG. 2.** (a) A bright-field scanning transmission electron microscope image of anatase $Ti_{0.90}Co_{0.10}O_2$ (001) thin film / $TiO_2$ (001) buffer layer / $LaAlO_3$ (100) substrate. (b) The energy-dispersive x-ray spectroscopy mappings for Co and Ti in (a). (c) Cross sectional image showing the location of images in (d) and (e). High resolution high-angle annular dark-field (HAADF)-STEM images (d) at $TiO_2$ / $LaAlO_3$ interface and (e) within $Ti_{0.90}Co_{0.10}O_2$ thin film [27].

**FIG. 3.** (a) Temperature dependence of resistivity $\rho_{xx}$ at 300 K. (b) The dependences of conductivity $\sigma_{xx}$ (solid circle), electron density $n$ (solid square), and mobility $\mu$ at 300K for anatase $Ti_{0.90}Co_{0.10}O_2$ thin films on the oxygen pressure during growth $P_{O2}$. The samples in (a) and (b) are indexed with the alphabets (A–L) [28].

**FIG. 4.** Magnetic field dependence of magnetization at 300 K for (a) anatase $Ti_{0.90}Co_{0.10}O_2$ thin films with different electron densities. Magnetic field was applied along the out-of-plane. Alphabetical sample index is the same as that in Fig. 3 [28]. (b) Magnetic field dependence of magnetization under out-of-plane (solid circle) and in-plane (open circle) magnetic fields at 300 K for anatase $Ti_{0.90}Co_{0.10}O_2$ thin films with different electron densities [27].

**FIG. 5.** (a) Absorption coefficient $\alpha$ and (b) magnetic circular dichroism (MCD) spectra for anatase $Ti_{1-x}Co_xO_2$ thin films ($x$ = 0, 0.1). The electron density at 300 K is displayed. The dashed lines correspond to the critical point energy in the band structure of anatase $TiO_2$. In (a), the data for $x$ = 0.1 is shifted vertically. In (b), the magnetic field was applied along the out-of-plane [36].

**FIG. 6.** (a) Magnetic field dependences of anomalous Hall resistivity $\rho_{AH}$ and magnetization for an anatase $Ti_{0.95}Co_{0.05}O_2$ thin film at 300 K. Inset shows magnetic field dependence of $\rho_H$ at 300 K [38]. (b) Magnetic field dependence of anomalous Hall conductivity $\sigma_{AH}$ at 300 K for anatase $Ti_{0.90}Co_{0.10}O_2$ thin films with different electron densities. Alphabetical sample index is the same as that in Fig. 3. The magnetic field was applied along the out-of-plane for all the measurements [28].

**FIG. 7.** The relation between anomalous Hall conductivity $\sigma_{AH}$ and conductivity $\sigma_{xx}$ for various ferromagnetic compounds modified from Ref. [41].

**FIG. 8.** (a) Temperature dependence of resistivity $\rho_{xx}$ for anatase $Ti_{0.90}Co_{0.10}O_2$ thin film channel during cooling with different gate voltages $V_G$. The inset shows schematic diagram of electric double layer transistor. (b) Temperature dependences of conductivity $\sigma_{xx}$, mobility $\mu$, and electron density $n$ at each $V_G$. An ionic liquid, N,N-diethyl-N-(2-methoxyethyl)-N-methylammonium bis- (trifluoromethylsulfonyl)-imide

(DEME-TFSI), was used as the electrolyte [27].

**FIG. 9.** (a) Magnetic field dependence of anomalous Hall conductivity $\sigma_{AH}$ at 300 K for anatase $Ti_{0.90}Co_{0.10}O_2$ thin film channel with different gate voltages $V_G$. $n$ at each $V_G$ is displayed. (b) The relation between $\sigma_{AH}$ and $n$ at 300 K. Black circles and triangles denote the data for two different electric double layer transistors (devices 1 and 2, respectively). Gray circles denote the data from nine $Ti_{0.90}Co_{0.10}O_2$ thin films, in which $n$ was varied by tuning the amount of the oxygen vacancy. The used electrolyte was N,N-diethyl-N-(2-methoxyethyl)-N-methylammonium bis- (trifluoromethylsulfonyl)-imide (DEME-TFSI) for device 1, and $CsClO_4$ solved in polyethylene oxide for device 2 [27].

**Fig. 10.** Electron density $n$ dependence of (a) incremental magnetization from paramagnetic state $M_{inc}$ and (b) anomalous Hall conductivity $\sigma_{AH}$ at 300K for anatase $Ti_{0.90}Co_{0.10}O_2$ thin films. Open symbols denote a superparamagnetic sample. Alphabetical sample index is the same as that in Fig. 3. In (b), data of rutile $Ti_{0.90}Co_{0.10}O_2$ are also shown [28].

Figure1

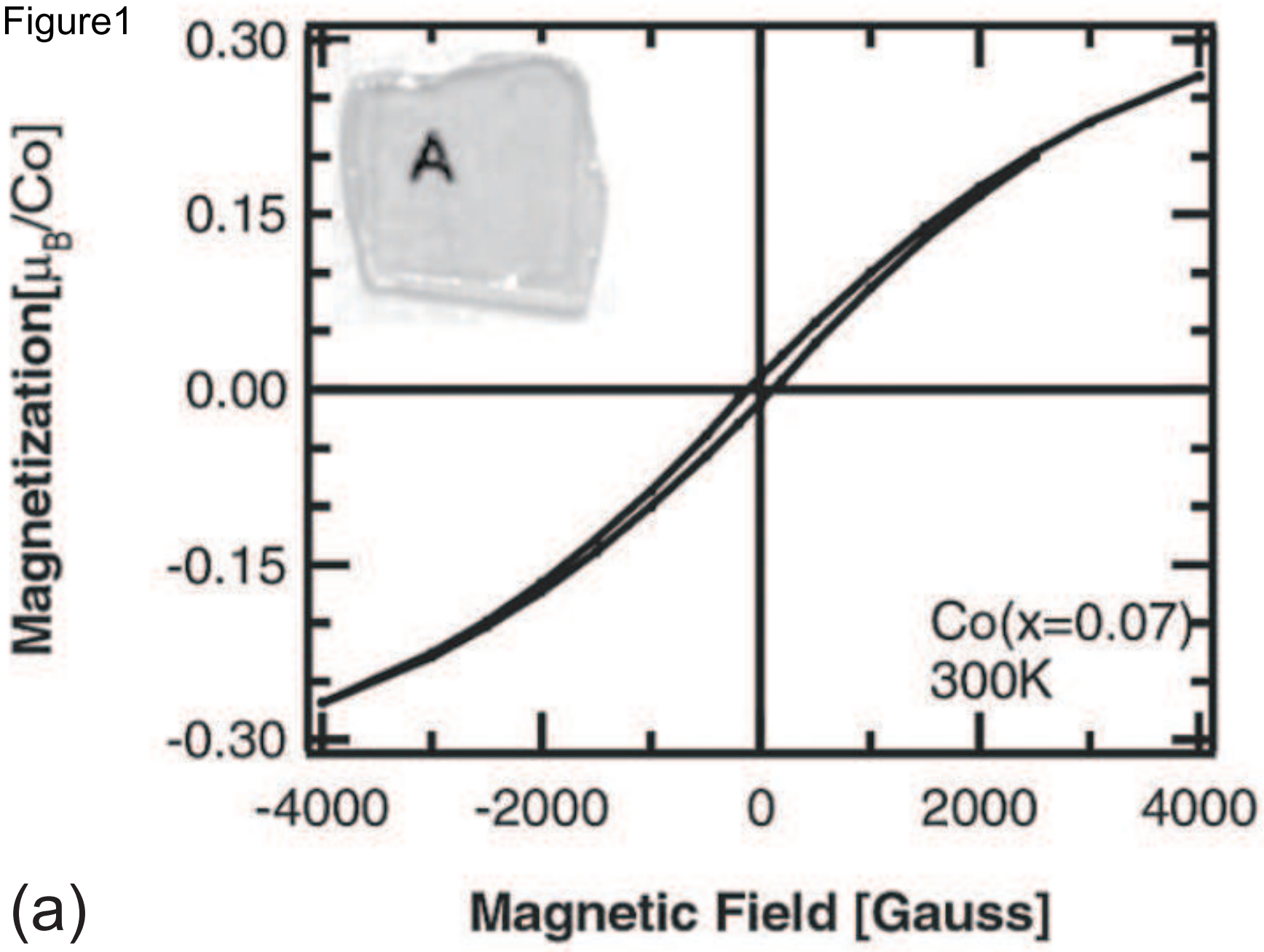

(a)

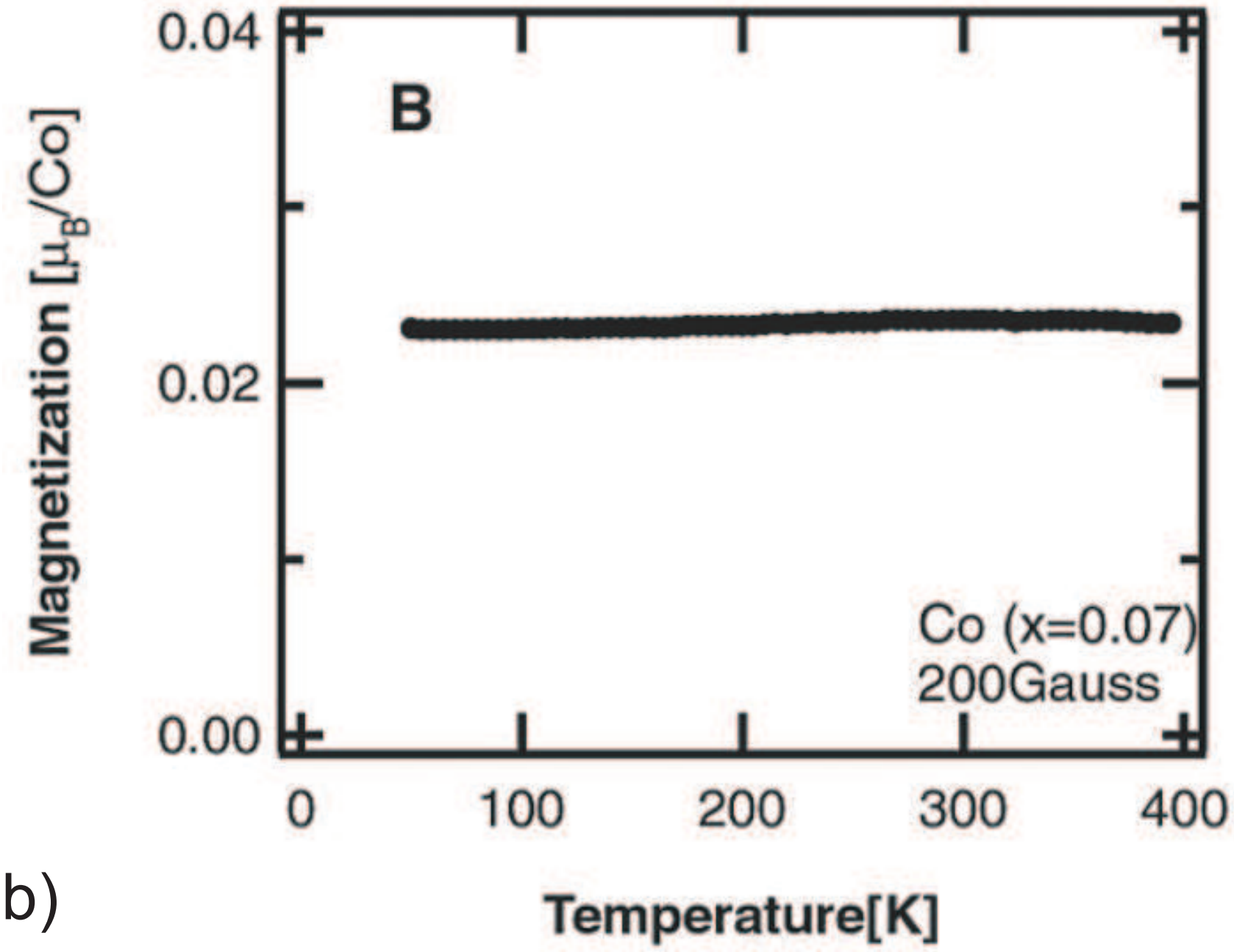

(b)

Fig. 1

Figure2

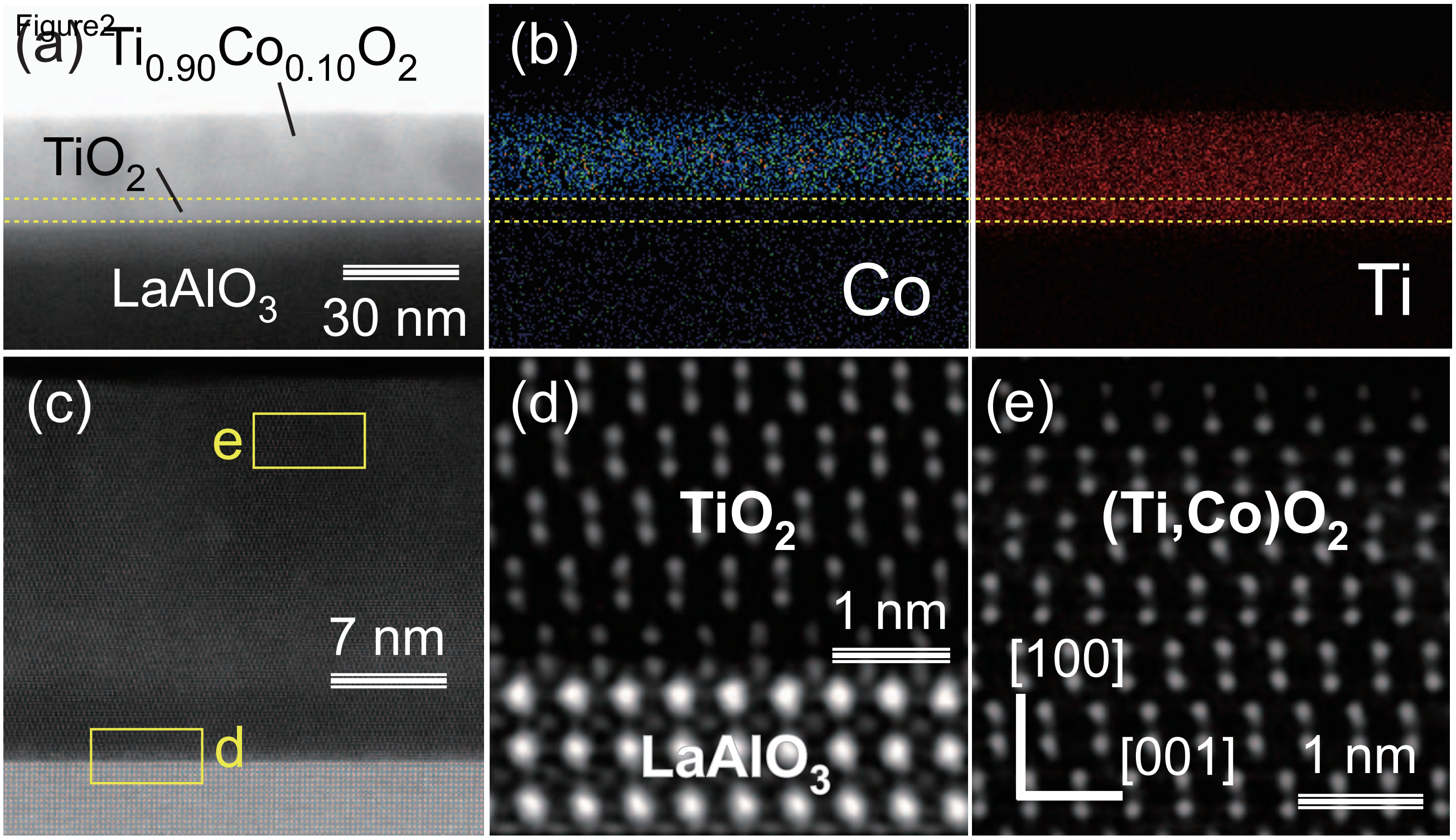

Fig. 2

Figure3

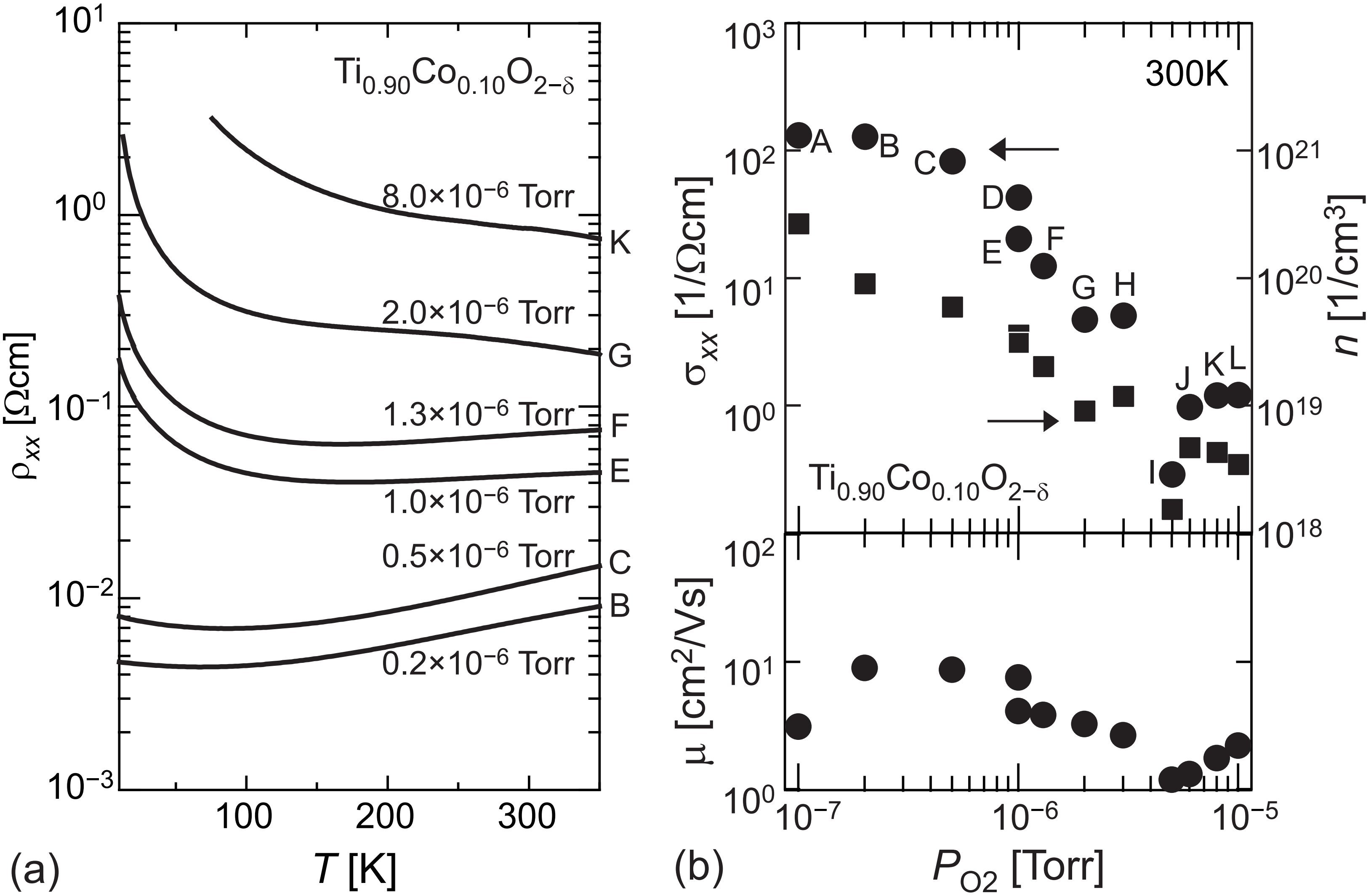

Fig. 3

Figure4

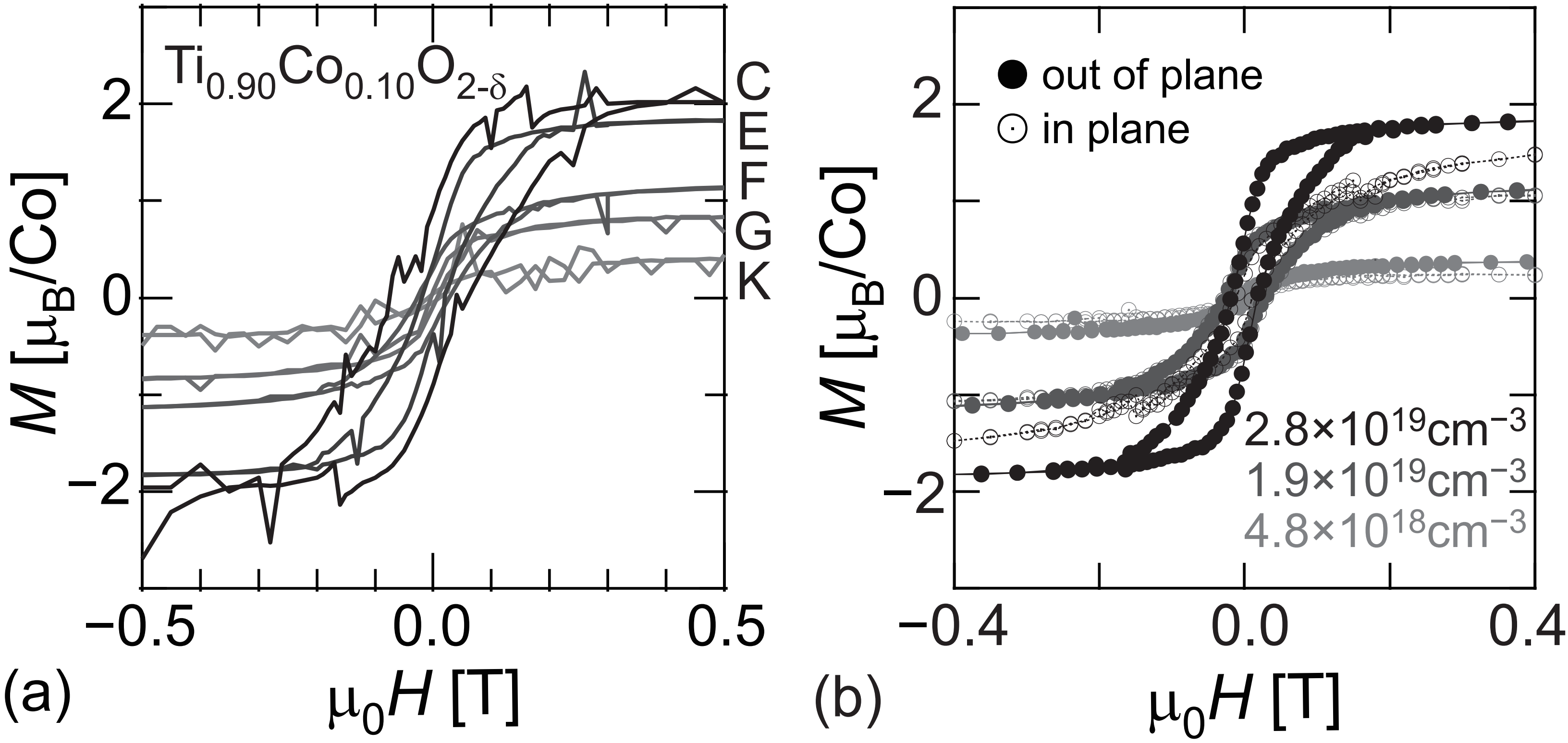

Fig. 4

Figure5

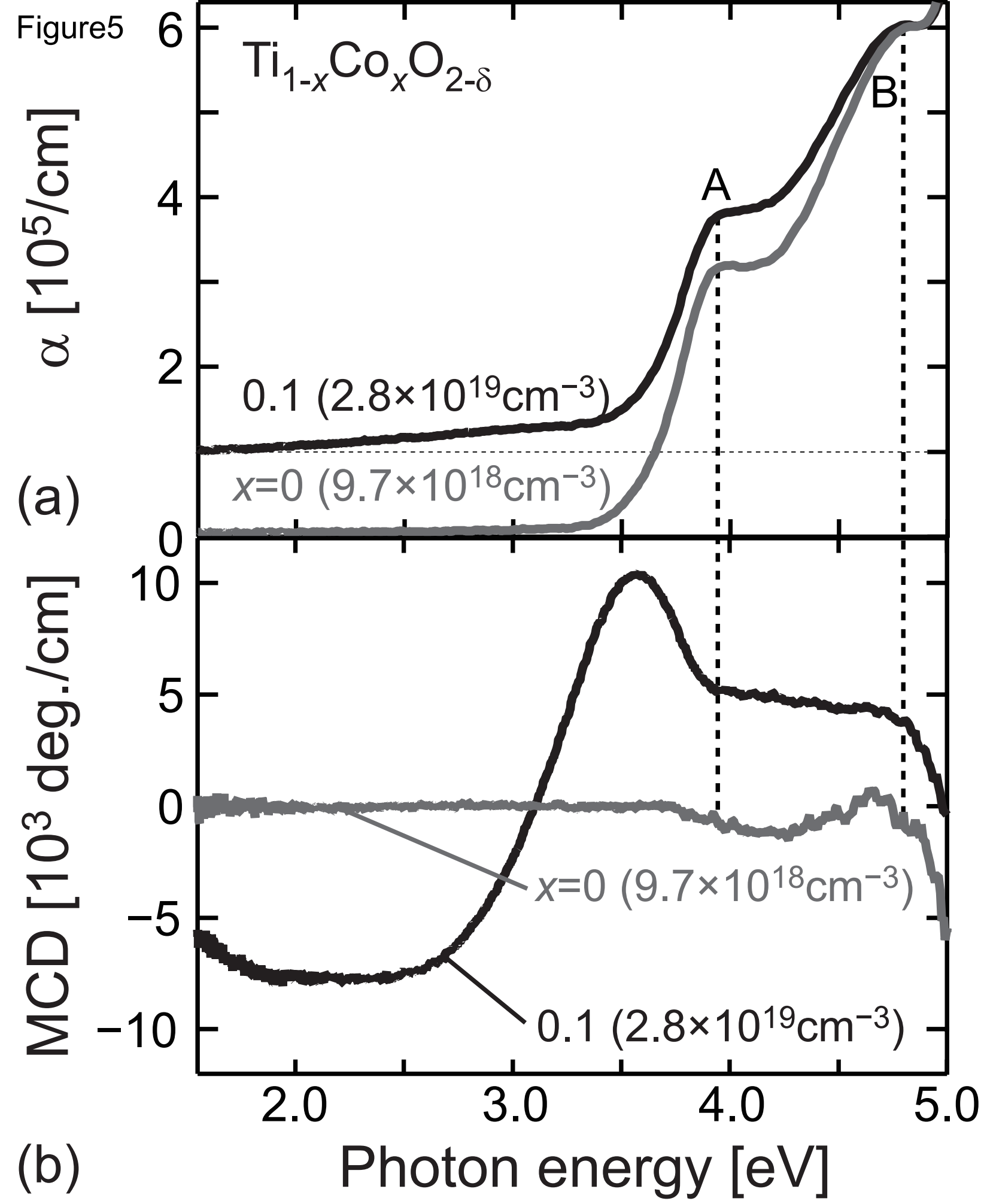

Fig. 5

Figure6

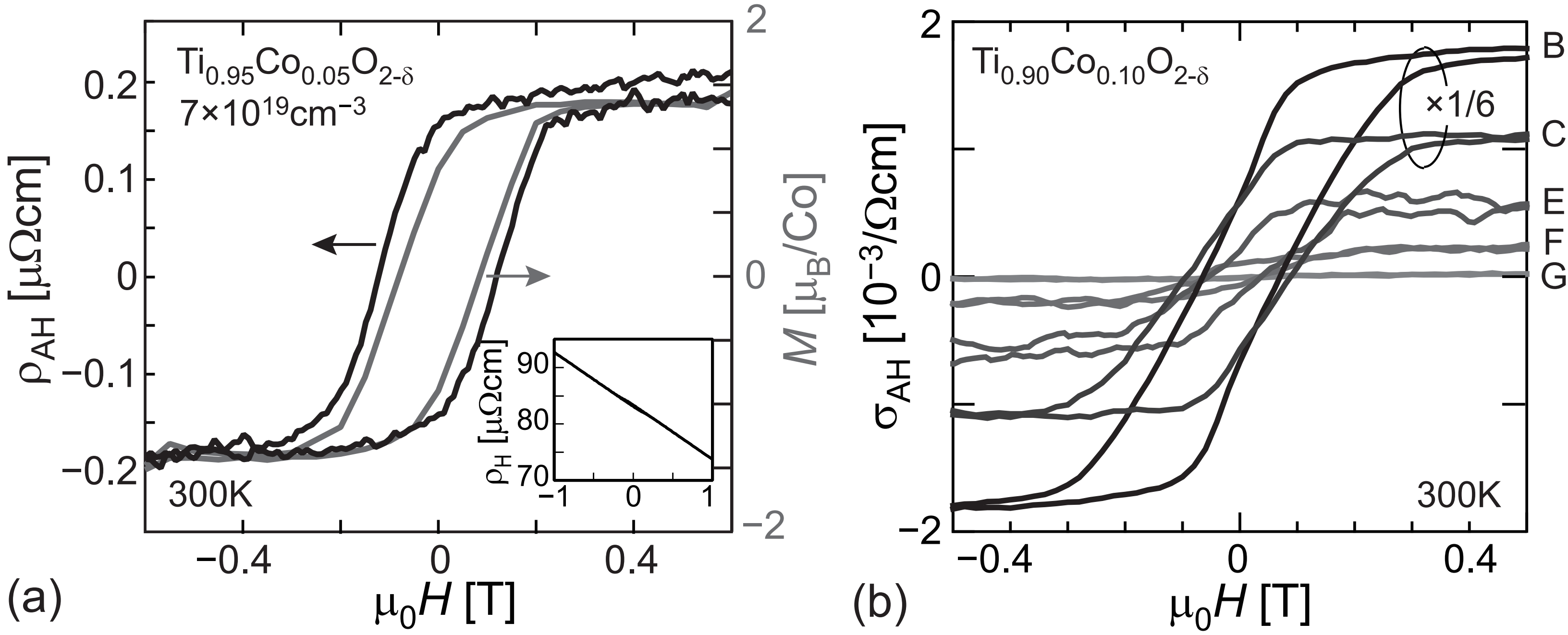

Fig. 6

Figure7

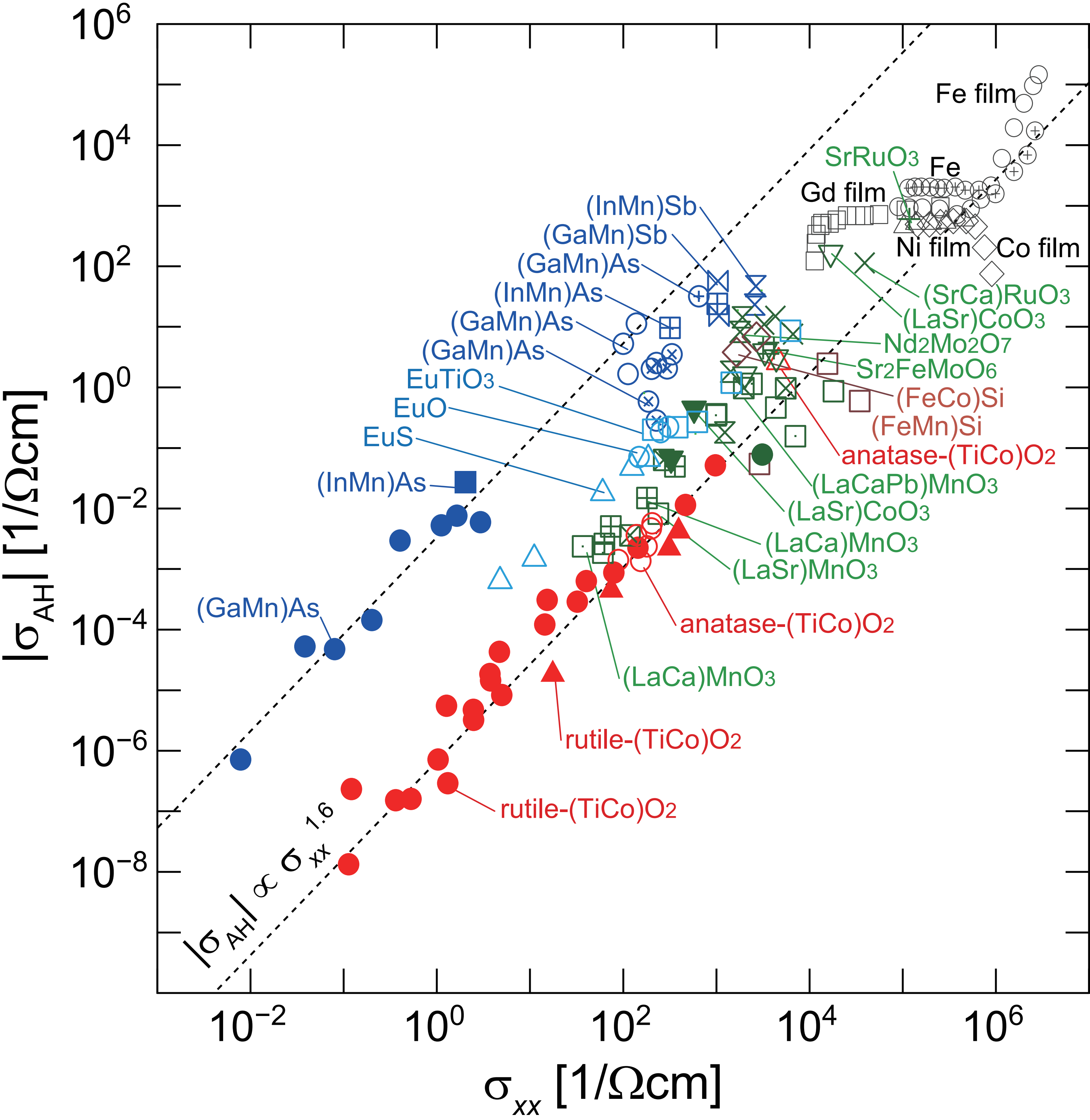

Fig. 7

Figure8

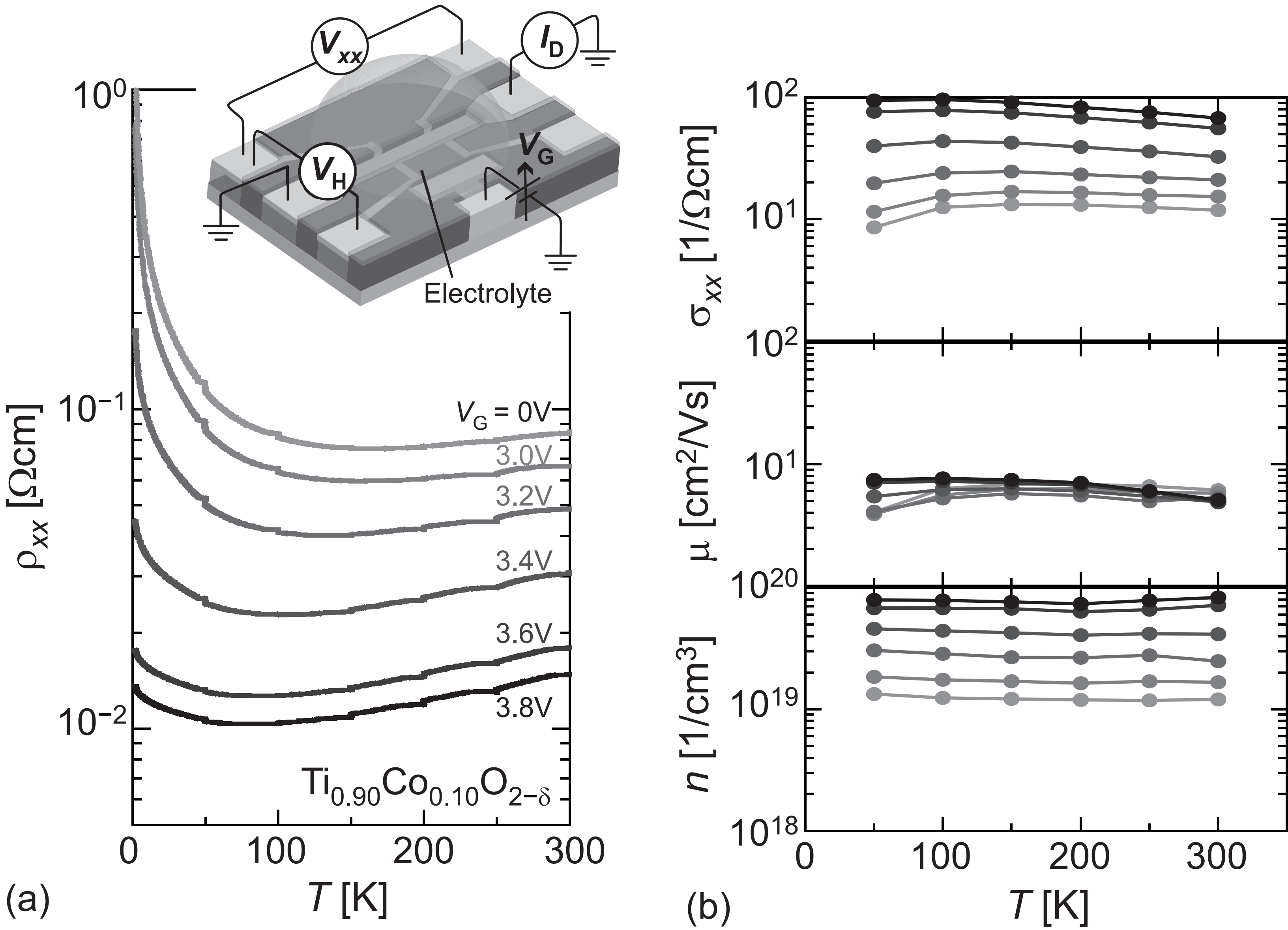

Fig. 8

Figure9

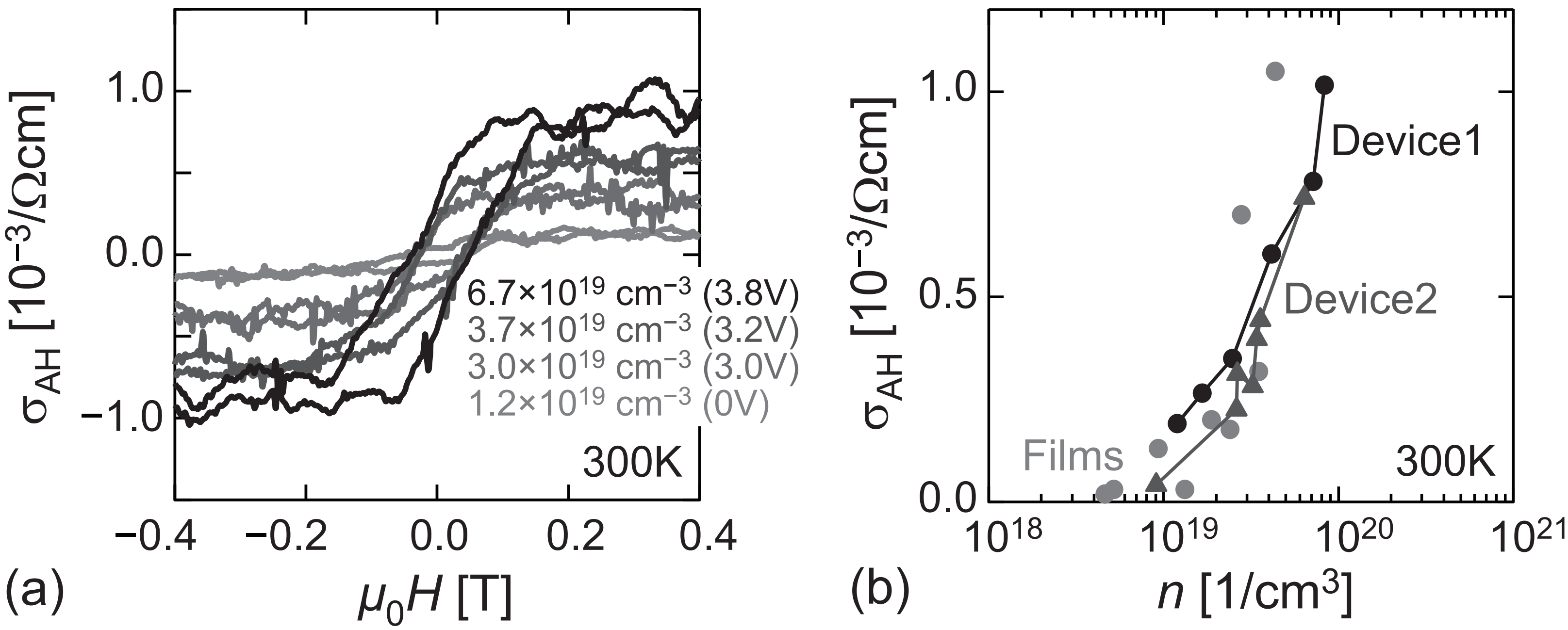

Fig. 9

Figure10

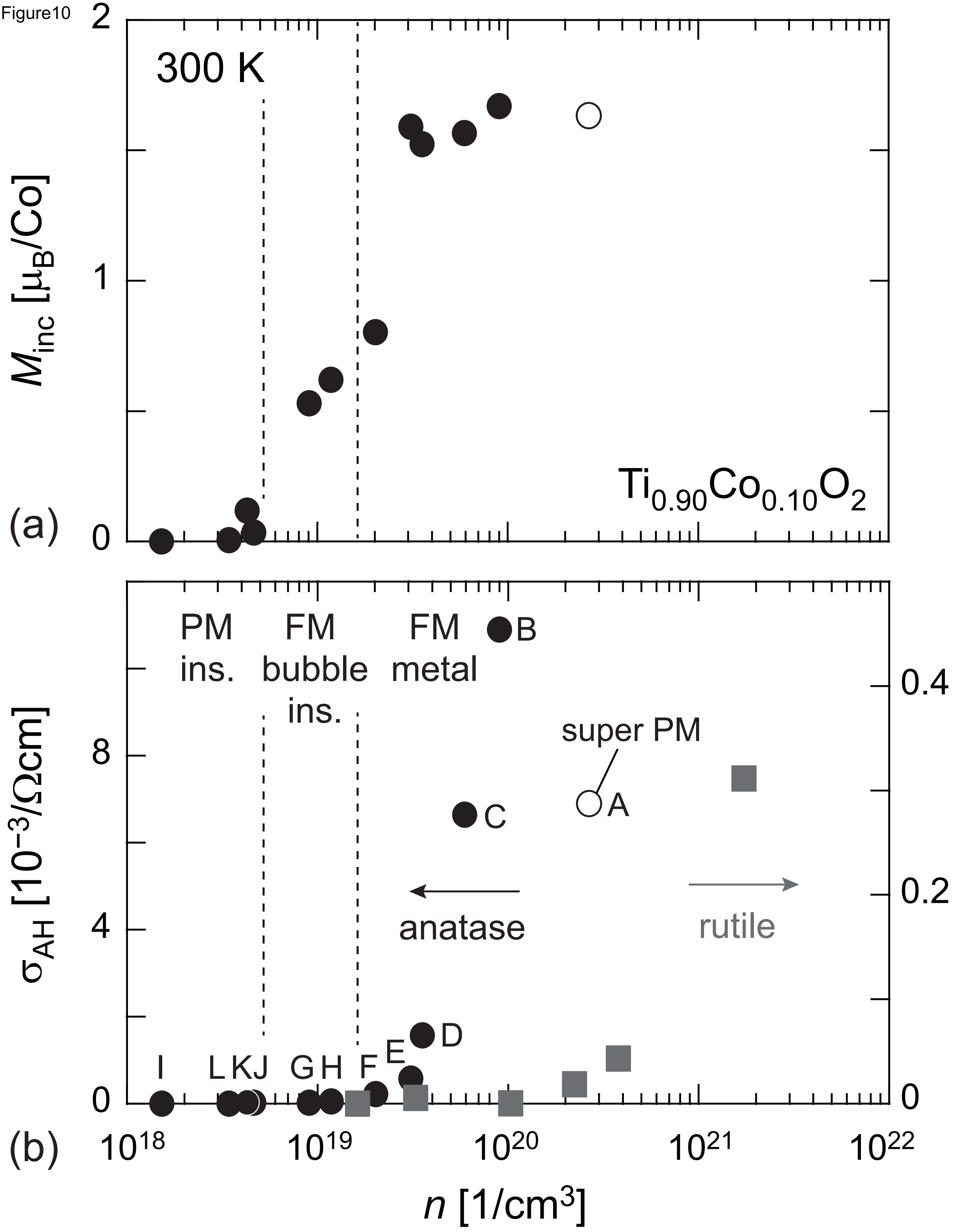

Fig. 10